\documentclass[prl,aps,twocolumn,10pt,superscriptaddress,notitlepage,longbibliography]{revtex4-2}

\usepackage[dvipsnames]{xcolor}
\usepackage[english]{babel} 
\usepackage{graphicx}
\usepackage{float}
\usepackage{braket}
\usepackage{epstopdf}
\usepackage{mathtools}
\usepackage{amsmath}
\usepackage{amssymb}
\usepackage{bbm,bm}
\usepackage[caption=false]{subfig}
\usepackage{pythonhighlight}
\usepackage{hyperref}
\usepackage{float}
\usepackage{bbold}
\usepackage[T1]{fontenc}

\usepackage{mathtools}
\usepackage{dsfont}
\usepackage{bm}
\graphicspath{{./Figures/}{./Figures_thesis/}}

\usepackage[markup=underlined]{changes}
\makeatletter
\@namedef{Changes@AuthorColor}{magenta}
\colorlet{Changes@Color}{magenta}
\makeatother
\usepackage{hyperref}
 \hypersetup{
     colorlinks=true,
     linkcolor=blue,
     filecolor=blue,
     citecolor = magenta,      
     urlcolor=red,
     }

\usepackage{braket}
\usepackage{xargs}
\newcommandx{\greencom}[2][1=]
{\todo[inline, color=green!40,#1]{#2}}
\newcommandx{\bluecom}[2][1=]
{\todo[inline, color=blue!40,#1]{#2}}
\newcommandx{\bluemargin}[2][1=]
{\todo[color=blue!40,#1]{#2}}

\usepackage{letltxmacro}
\LetLtxMacro{\ORIGselectlanguage}{\selectlanguage}
\makeatletter
\DeclareRobustCommand{\selectlanguage}[1]{%
  \@ifundefined{alias@\string#1}
    {\ORIGselectlanguage{#1}}
    {\begingroup\edef\x{\endgroup
       \noexpand\ORIGselectlanguage{\@nameuse{alias@#1}}}\x}%
}
\newcommand{\definelanguagealias}[2]{%
  \@namedef{alias@#1}{#2}%
}

\makeatother

\definelanguagealias{en}{english}
\definelanguagealias{EN}{english}
\definelanguagealias{eng}{english}
\definelanguagealias{de}{ngerman}

\newcommand{\ii}{\mathrm{i}}

\usepackage{setspace}
\usepackage{xr}
\makeatletter

\newcommand*{\addFileDependency}[1]{% argument=file name and extension
\typeout{(#1)}% latexmk will find this if $recorder=0
\@addtofilelist{#1}
\IfFileExists{#1}{}{\typeout{No file #1.}}
}\makeatother

\newcommand*{\myexternaldocument}[1]{%
\externaldocument{#1}%
\addFileDependency{#1.tex}%
\addFileDependency{#1.aux}%
}
\myexternaldocument{SI}
\begin{document}

\title{Floquet engineering the quantum Rabi model in the ultrastrong coupling regime}

%%%%%%%%%%%%%%%%%%%%%%%%%%%%%%%%%%%%%%%%%%%%%%%%%%%%%%%%
% Authors: Kamran, Steve
%%%%%%%%%%%%%%%%%%%%%%%%%%%%%%%%%%%%%%%%%%%%%%%%%%%%%%%%
\author{Kamran~Akbari}
\email[]{kamran.akbari@queensu.ca}
\affiliation{Department of Physics, Engineering Physics and Astronomy, Queen's University, Kingston ON K7L 3N6, Canada}
%------------------------------------------------------------------
\author{Franco Nori}
\affiliation{Theoretical Quantum Physics Laboratory, Cluster for Pioneering Research, RIKEN,  Wakoshi, Saitama 351-0198, Japan}
\affiliation{Quantum Computing Center, RiKEN, Wakoshi, Saitama, 351-0198, Japan}
\affiliation{Physics Department, The University of Michigan, Ann Arbor, Michigan 48109-1040, USA}
%------------------------------------------------------------------
\author{Stephen Hughes}
%\email[]{shughes@queensu.ca}
\affiliation{Department of Physics, Engineering Physics and Astronomy, Queen's University, Kingston ON K7L 3N6, Canada}
%------------------------------------------------------------------
\date{\today}

%%%%%%%%%%%%%%%%%%%%%%%%%%%%%%%%%%%%%%%%%%%%%%%%%%%%%%%%
% section: abstract
%%%%%%%%%%%%%%%%%%%%%%%%%%%%%%%%%%%%%%%%%%%%%%%%%%%%%%%%
\begin{abstract} 
We study the quantum Rabi model
for a two-level system coupled to
a quantized cavity mode
under periodic modulation of the
cavity-dipole coupling in the 
ultrastrong coupling regime, leading to rich Floquet states. As an application of the theory, we show how purely mechanical driving can 
produce real photons,
depending on the 
strength and frequency of the 
periodic coupling rate.
\end{abstract}

\maketitle

Since the early 
observations with
intersubband polaritons~\cite{Anappara_Signatures_2009,Zaks2011},
the ultrastrong coupling (USC) regime of light-matter interactions has 
emerged as  a fascinating and  
distinctive phenomenon in quantum optics, particularly within the realm of cavity quantum electrodynamics 
(QED)~\cite{FriskKockum_Ultrastrong_2019,Forn-Diaz_Ultrastrong_2019}. 
A particularly intriguing facet of 
USC
 is the occupation of virtual excitations (e.g., photons), even in the ground state
 of the dressed system, 
 which is a  
consequence of counter-rotating wave effects and breaking $U(1)$ symmetry~\cite{DeLiberato_Virtual_2017,FriskKockum_Ultrastrong_2019}. This  raises the question of whether it is possible to convert  virtual photons into real 
ones~\cite{FriskKockum_Deterministic_2017,FriskKockum_Ultrastrong_2019},
which  requires input energy, e.g., 
coherent/incoherent excitation~\cite{Werlang_Rabi_2008,Cirio_Amplified_2017}, 
to introduce time-dependent characteristics into the system, nonadiabatically~\cite{Ciuti_Quantum_2005,Nation_Colloquium:Stimulating_2012}.

Although virtual excitations 
%and particularly virtual photons, 
are not detectable, ideas have been proposed to 
release 
them as real excitations~\cite{Lolli_Ancillary_2015,Cirio_Amplified_2017,Ridolfo_Probing_2021,
Ciuti_Quantum_2005,DeLiberato_Quantum_2007,Takashima_Nonstationary_2008,Werlang_Rabi_2008,Dodonov_Photon_2008,Dodonov_Photon_2009,Beaudoin_Dissipation_2011,DeLiberato_Extracavity_2009,Carusotto_Back-reaction_2012,Garziano_Switching_2013,Shapiro_Dynamical_2015,Gunter_Sub-cycle_2009,Ridolfo_Optical_2011,Huang_Photon_2014,Cirio_Ground_2016,Minganti_Phonon_2023,Mercurio_Flying_2022}.
These 
methods typically 
involve modulating one of the system parameters~\cite{Ciuti_Quantum_2005,DeLiberato_Quantum_2007,Takashima_Nonstationary_2008,Werlang_Rabi_2008,Dodonov_Photon_2008,Dodonov_Photon_2009,Beaudoin_Dissipation_2011,DeLiberato_Extracavity_2009,Carusotto_Back-reaction_2012,Garziano_Switching_2013,Shapiro_Dynamical_2015,Minganti_Phonon_2023,Mercurio_Flying_2022}, e.g.,
time-modulation of the 
%intersubband 
Rabi 
frequency~\cite{Ciuti_Quantum_2005,DeLiberato_Quantum_2007,Ciuti_Quantum_2005,DeLiberato_Quantum_2007}, 
using flying atoms~\cite{Mercurio_Flying_2022},
and exploiting phonon pumping~\cite{Minganti_Phonon_2023,Macri_Nonperturbative_2018}.
Nevertheless, the first theory proposals~\cite{Johansson_Dynamical_2009,Johansson_Dynamical_2010,Johansson_Nonclassical_2013} that paved the way to the successful experiments~\cite{Wilson_Observation_2011,Dalvit_Shaking_2011} (particularly, in circuit QED) were mostly associated with the dynamical Casimir effect~\cite{Macri_Nonperturbative_2018, DiStefano_Interaction_2019,Settineri_Conversion_2019,Qin_Emission_2019}.

The quantum Rabi model (QRM), where
a two-level system (TLS) is coupled to a
single quantized cavity mode,
is the key model in cavity-QED.
In 
%the 
USC,
%regime, 
there is a 
 prominent role played by counter-rotating wave effects and ponderomotive energies,
 and the 
 %, which are often overshadowed and negligible in other coupling regimes.
% In contrast,
%For example, 
%the 
Jaynes-Cummings model, a cornerstone rotating-wave model
for explaining 
%cavity and circuit QED experiments,
%under 
weak and strong coupling effects,
%in cavity-QED, 
is no longer 
valid~\cite{DeBernardis_Breakdown_2018,Salmon_Gauge-independent_2020,Akbari_Generalized_2023}. Instead,
one must consider  the
joint atom-cavity 
%systems hybridize into
 dressed states~\cite{FriskKockum_Ultrastrong_2019, Settineri_Dissipation_2018,Salmon_Gauge-independent_2020,Akbari_Generalized_2023},
where even the ground state is an entangled state of photons and matter.
 Moreover,
 it is essential to uphold the gauge-invariance principle when dealing with truncated 
 %quantum 
 matter 
models~\cite{DeBernardis_Breakdown_2018,DiStefano_Resolution_2019,Taylor_Resolution_2020,Taylor2022, Settineri_Gauge_2021,Salmon_Gauge-independent_2020,Akbari_Generalized_2023,Gustin_Gauge-invariant_2023}.

However, for driven 
cavity-QED systems, the QRM can also fail,
since 
%in USC, 
the strong hybridization 
of the bare subsystems, 
demands a nonperturbative treatment~\cite{Settineri_Dissipation_2018,Salmon_Gauge-independent_2020} before and after driving.
Often, periodic driving 
is 
%then 
considered
as a weak perturbation 
that induces 
transitions 
%perturbations 
between the
(pre-driving) hybrid states~\cite{Salmon_Gauge-independent_2020,Macri_Spontaneous_2022}. 
Yet, when the
strength of the driving amplitude
%of a time-dependent dressing
is also significant, 
%then
the dressed (joint) light-matter states of the entire system 
transforms into a Floquet picture, 
an important theoretical framework for %transcending the rotating-wave approximation in
understanding periodically driven systems~\cite{Floquet_FloquetTheory_1883,Grifoni_Driven_1998}.
Apart from fundamental interest,
 Floquet theory is a powerful tool for engineering quantum systems~\cite{Grifoni_Driven_1998,Iadecola_Floquet_2015,Oka_Floquet_2019,Diermann_Floquet-state_2019,Weitenberg_Tailoring_2021,Bandyopadhyay_Floquet_2022,Petiziol_Cavity-Based_2022,Liu_Floquet_2023,Castro_Floquet_2023,Zhan_Floquet_2023,Geier_Floquet_2021,Bai_Floquet_2023} and reservoirs~\cite{Schnell_Dissipative_2023,Kumar_Floquet_2022},
 and has been  used for
 describing 
 photon-assisted quantum  tunneling/transport~\cite{Tien_Multiphoton_1963,Grossmann_Coherent_1991,Grifoni_Driven_1998,Platero_Photon-assisted_2004,Eckardt_Analog_2005,Shibata_Photon-assisted_2012,Coish_Entangled_2009}. To the best of our knowledge, it has not been utilized for the USC regime.

In this work, we
describe how one can {\it Floquet engineer the
QRM}, by applying 
nonperturbative periodic oscillations to the 
TLS-cavity coupling rate.
The hybrid system 
%energy 
states evolve nonadiabatically into  Floquet quasienergy states, forming new transitions via the newly-introduced anticrossings in the Floquet picture.
%This mechanism 
%is made possible through a
%periodic oscillation of the 
%atom-cavity coupling rate,
This type of periodic modulation 
%which 
can 
%also 
connect to various experimentally accessible regimes, such as
the dynamical Casimir effect~\cite{Macri_Nonperturbative_2018}, surface acoustic waves in semiconductors~\cite{2019_Delsing_2019,Iikawa_Optical_2016}, and optomechanical interactions~\cite{Cirio_Amplified_2017},
including molecular optomechanics~\cite{Roelli2015,Schmidt2016,Dezfouli2019}.
Our significant findings include: (i) 
%the prediction of a   
a double-field (photon plus mechanical oscillation)-assisted splitting of the QRM states due to the renormalization of the time-independent energy states, (ii) production of real photons and TLS excitations from vacuum, (iii) higher-order nonlinear quantum processes that are effective only in the USC regime.

%%%%%%%%%%%%%%%
% \section{Time-dependent quantum Rabi model in USC} 
%%%%%%%%%%%%%%%

\begin{figure}[t]
    \centering        %\includegraphics[width=.95\linewidth]{Figures/Fig1_V1.pdf}
    \includegraphics[width=.95\linewidth]{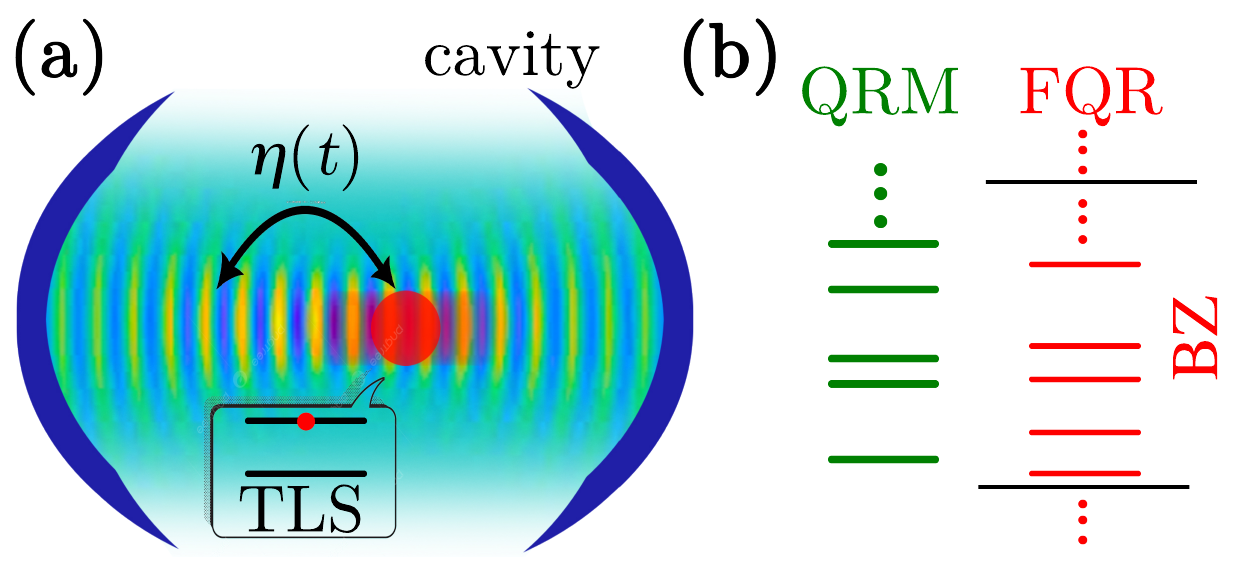}
        \caption[]{
        %\textbf{Cavity-QED with time-dependent coupling.}
    (a) Schematic of a TLS with a mechanical vibration inside a cavity with a dominant single mode. (b) Without the mechanical vibration, the system is identified by the usual (time-independent) QRM Hamiltonian with $N_j$ dressed states (left); after the periodic vibration is turned on, the Floquet quasi-energies and states (FQR) govern the system (right) transitions, with $N_j$ of them in one Brillouin zone (BZ).
   }
    \label{Fig1}
\end{figure}

We begin with  the \textit{time-dependent}
QRM 
Hamiltonian,
\begin{equation}
 {\mathcal{H}}_\mathrm{FQR}(t)=\omega_\mathrm{c}{a}^\dagger{a} 
+ \frac{\omega_{a}}{2}\left \{\sigma_z\cos[ c(t)]
+\sigma_y\sin[c(t)] \right \}
,
\label{HC_FQR}
\end{equation}
in the Coulomb gauge~\cite{DiStefano_Resolution_2019} ($\hbar{=}1$), where $\omega_c$ ($\omega_a$) is the cavity (TLS)
transition frequency, $a$ ($a^\dagger$) is the cavity photon annihilation (creation) operator, $\sigma_{i}$ are the TLS Pauli operators, and 
$c(t) = 2({a}+{a}^\dagger)\eta(t)$.
The normalized TLS-cavity  coupling rate is %given by 
$\eta(t)=\eta_0+\eta_{{M}}\sin(\omega_{ M}t)$, 
 where $\eta_0 \equiv g/\omega_c$  is the 
normalized time-independent  coupling rate ($g$ is the atom-cavity coupling rate), and
$\eta_{M}$ is the amplitude of the time-dependent coupling,
with $\omega_{M}$ the frequency of the 
%externally 
driven time-dependent oscillation. 
The calligraphic notation of the Hamiltonian indicates that the gauge-fixed Hamiltonian is used for the {\it truncated} matter Hilbert space~\cite{DiStefano_Resolution_2019,Salmon_Gauge-independent_2020,Gustin_Gauge-invariant_2023}. 
Note that when
$\eta(t)\to\eta$, then Eq.~\eqref{HC_FQR} fully recovers
previous 
 time-independent (and  gauge invariant) models~\cite{DiStefano_Resolution_2019,Salmon_Gauge-independent_2020,Akbari_Generalized_2023}.
 Figure~\ref{Fig1} shows a schematic
of our time-dependent QRM.
%cavity-QED scheme.

Due to the periodicity of the time-dependent coupling, with period $T=2\pi/\omega_{M}$, the Hamiltonian is also periodic: $\mathcal{H}_{\rm FQR}(t) =\mathcal{H}_{\rm FQR}(t+T)$, which can be expanded as a Fourier series 
${\mathcal{H}}_\mathrm{FQR}(t)=\sum_{m\in\mathbb{Z}}\mathcal{H}_m\,\mathrm{e}^{\ii m\omega_{ M}t}$, with
\begin{equation}
\begin{split}\displaystyle
 \mathcal{H}_m
&=\omega_\mathrm{c}{a}^\dagger{a} \,\delta_{m0}+\dfrac{\omega_{a}}{2}
%\lcb
\left \{
\dfrac{\sigma_z-\ii\sigma_y}{2}\,\mathrm{e}^{\ii2({a}+{a}^\dagger)\eta_0} \right.
\\
& 
\hspace{-0.5cm}\left.
+(-1)^m\dfrac{\sigma_z+\ii\sigma_y}{2}\,\mathrm{e}^{-\ii2({a}+{a}^\dagger)\eta_0} 
\right \}
J_m[2({a}+{a}^\dagger)\eta_{M}],
\end{split}
\label{H_m}
\end{equation}
where $J_m$ is the Bessel function of the first kind of order $m$, and we have
used the Anger-Jacobi expansion~\cite{Abramowitz_Handbook_1965} of Eq.~\eqref{HC_FQR}.
When
$\eta_{M} \rightarrow 0$, then
$J_0(0)=1$ and $J_{m\gtrless0}(0)=0$, and we recover the time-independent QRM
 Hamiltonian. 
 In practice, we must also
 truncate 
 %the number of 
 %harmonics by 
 $\lvert m\rvert\leq m_{\rm max}$.
 % where  $m$ counts the number of the harmonic process (exchange of energy packets of $\omega_{M}$).
 % %, i.e., exchange of energy packets of $\omega_{\rm M}$, absorption %($m>0$) and emission ($m<0$).
Note also that 
%the full Hamiltonian is
$\mathcal{H}_m$ separates into a time-independent part for $m=0$ (including 
a shift due to $\eta_{M}\neq0$), 
%as well as
and a time-dependent interaction (for $m \neq 0$).
Thus, while 
%It is interesting to point out that due to the nonzero presence of 
$\eta_{M}$ is  related to the %emergence of 
time-dependent light-matter interaction, there is a static contribution 
from the $J_0$ term. 
While Hamiltonian~\eqref{HC_FQR}
accounts for the static dressing
via photon-matter interactions, 
 Eq.~\eqref{H_m} {\it dresses} the entire
cavity-QED system with periodic mechanical
oscillations.
 Similar expressions are widely considered for single quantum systems, including field-driven TLSs~\cite{Ashhab_Two-level_2007}. 
% around the region of avoided level 
% crossings~\cite{Ashhab_Two-level_2007}.

% Figure 1 ---------------------------------

For numerical calculations,
the time-independent 
%Hamiltonian, 
$\mathcal{H}_0$, is first diagonalized with the eigenbasis $\{E_j,\vert j\rangle\}_{j=0}^{N_j-1}$, where $N_j$ is the number 
of truncated AC-shifted  
QR-dressed states,
obtained from
%In other words, 
%from 
$\mathcal{H}_{0}\vert j\rangle=E_j\vert j\rangle$, which satisfies the conditions $\langle j\vert j'\rangle=\delta_{jj'}$ and $\sum_{jj'}\vert j\rangle\langle j'\vert={\bf 1}$; 
here
 $E_j$ ($\lvert j\rangle$) are 
shifted QRM eigenenergies (eigenstates) renormalized by the presence of the nonzero $\eta_{\rm M}$,
since  the time-independent portion of the 
%system 
Hamiltonian is $\mathcal{H}_0$,
and not $\mathcal{H}_{\rm QRM}=\omega_ca^\dagger a+(\omega_a/2)\{\sigma_z\cos[c(0)]+\sigma_y\sin[c(0)]\}$. 
This 
also 
%helps to 
ensures that we use the
correct static states of the joint light-matter system
in the presence of driving.

Solving the time-dependent Schrödinger equation, $\ii\partial_t\vert\psi(t)\rangle=\mathcal{H}_{\rm FQR}(t)\vert\psi(t)\rangle$, yields 
$\lvert\psi_\alpha(t)\rangle=\mathrm{e}^{-\ii\varepsilon_\alpha t}\lvert \alpha(t)\rangle$,
where $\varepsilon_\alpha$ 
is
the Floquet 
%(time-independent) 
quasienergy~\cite{Nikishov_Quantum_1964}, and the Floquet mode $\lvert \alpha(t)\rangle$ is  
$T$-periodic~\cite{Floquet_FloquetTheory_1883,Chicone_Ordinary_2006}.
%(analogous to Bloch modes in periodic lattices)~\cite{Floquet_FloquetTheory_1883,Chicone_Ordinary_2006}. 
The Floquet Hamiltonian, 
%is
%then defined via 
$\mathcal{H}_{\rm F}(t)\equiv  \mathcal{H}_{\rm FQR}(t)-\ii\partial/\partial t$, is associated with a set of quasienergies $\{\varepsilon_\alpha\}$.
 %Although time-dependent, 
 The 
 %set of 
 Floquet states, $\{\vert \psi_\alpha(t)\rangle\}$, form a complete basis for any value of $t$, 
 thus $\lvert\psi(t)\rangle=\sum_\alpha c_\alpha\lvert\psi_\alpha(t)\rangle$, where
 $c_\alpha=\langle\alpha\vert \psi(0)\rangle$, with $\vert \alpha\rangle\equiv\vert \alpha(0)\rangle$.
Transition resonances 
%of the system 
%can now 
occur at differences between Floquet energies~\cite{Shirley_Solution_1965}.
To compute the Floquet modes, we use a Fourier series expansion of
$\lvert \alpha(t)\rangle=\sum_{l\in\mathbb{Z}}\mathrm{e}^{\ii l\omega_{\rm M}t}\lvert\alpha_l\rangle$, where the Fourier coefficient states $\lvert\alpha_l\rangle$ are %called the 
{\it Floquet sidebands}.
%Correspondingly, 
There
%and there are
%According to the Floquet-Fourier approach, 
are $N_j$ quasienergies confined within a $[-\omega_{ M}/2,\omega_{M}/2]$ energy range 
(first BZ),
%Brillouin zone (BZ), 
%and they are 
associated with $N_j$ linearly independent Floquet modes~\cite{Akbari_Floquet_2023_SI}.

% Figure 2 ---------------------------------
\begin{figure}[t]
    \centering
        \includegraphics[width=.98\linewidth]{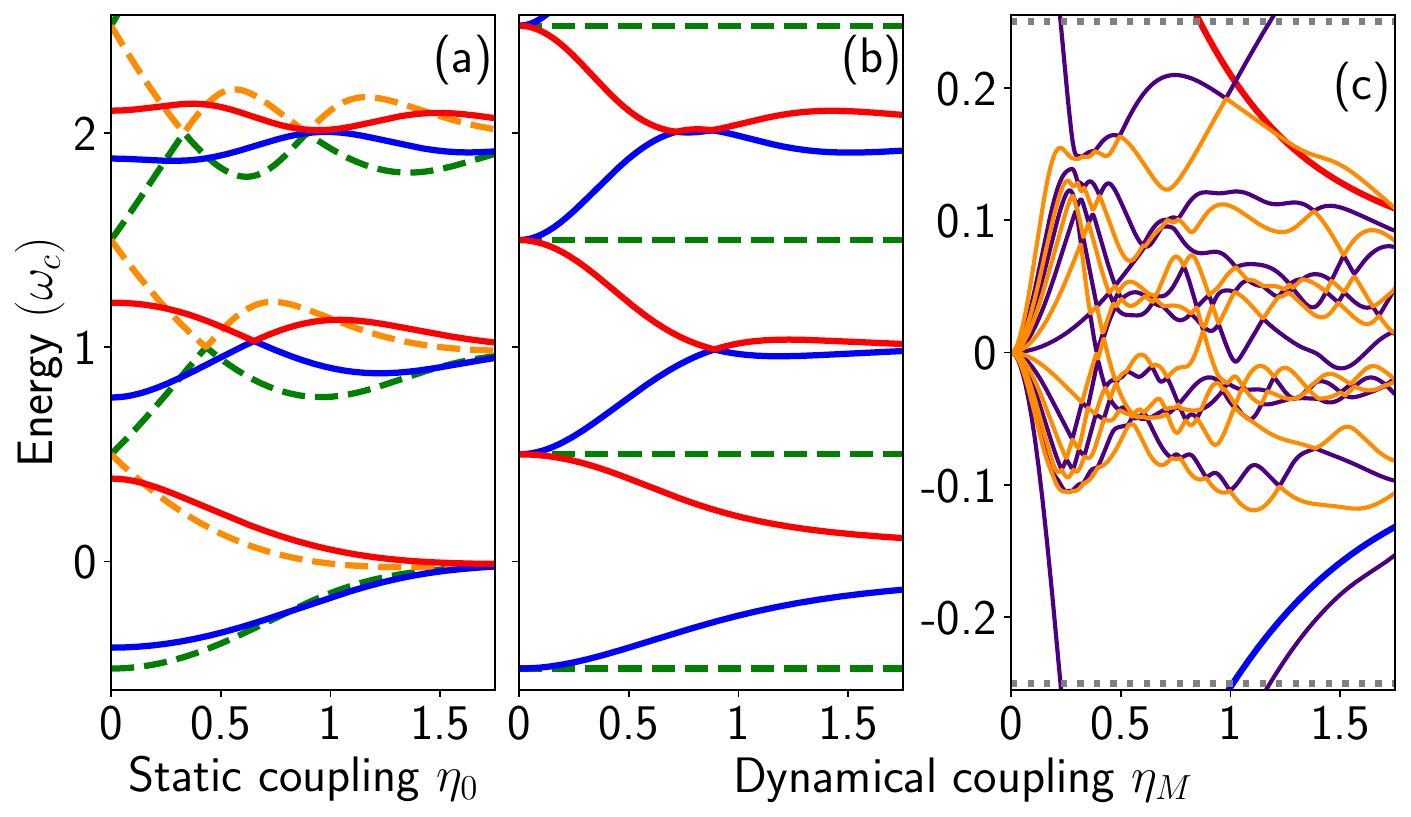}
        \caption[]{
        %\textbf{System energy basis.} Panel 
        (a) Eigenenergies obtained from 
        %from finding the eigenvalues of 
        Eq.~\eqref{H_m} for $m=0$ 
        %($\mathcal{H}_0$) 
        with $\eta_{ M}=0$ (QRM, dashed curved), and  for $\eta_{ M} =0.5$ (solid lines), versus $\eta_{0}$.
        (b) Eigenenergies obtained from  Eq.~\eqref{H_m} again for $m=0$ (i.e., static part), but  
        now as a function of finite $\eta_{M}$ 
        (solid lines), with $\eta_0=0$, as well as the QRM eigenenergies  (dashed lines). 
        %to demonstrate the static renormalization of the eigenenergies.
         (c) Floquet quasienergies in the $1^\mathrm{st}\,$BZ for $\eta_0=0$, versus $\eta_M$. Thin solid (with every other indigo and dark orange color to show anticrossings) lines represent different Floquet quasienergies.  Thicker  blue (lower) and red (upper) curves are the first two shifted eigenenergies from panel (b). The parameters used are: $\omega_M=0.5\omega_c$, $\omega_a = \omega_c$, $N_j=16$, $m_{\rm max}=l_{\rm max} = 20$ (see text). 
   }
    \label{Fig2}
\end{figure}
%-------------------------------------------

%--- DISCUSSION OF FIG2 ---
In Fig.~\ref{Fig2}(a), we plot the eigenenergies versus $\eta_0$, from  the 
{\it time-independent} part of the total Hamiltonian in Eq.~\eqref{HC_FQR}, i.e.,  using  $\mathcal{H}_0$ from Eq.~\eqref{H_m}, for a fixed value of $\eta_{ M}=0.5$ (solid curves),
and show how these compare with the
standard QRM (dashed curves). This
demonstrates 
how time-modulation 
introduces an effective static (DC) dressing of the 
QRM eigenenergies.
% the QRM eigenenergies and the initial splitting and anticrossings of the levels due to the atom-photon dressing. 
Also, we note from the form of the solid curves, that the levels are initially split, i.e., at
$\eta_0=0$, with more significant splittings
for higher energy levels.
% when there exists a coupling even though time-dependent which emphasizes the onset of dressing by the mechanical oscillation. 
As $\eta_0$ increases, we then enter a regime of \emph{double dressing}.
% at time $t=0^+$ \sh{presumably, since these are quasenergies, this must be $t\gg =0^+$
% or simply $t \gg T$, but it should not be sudden, just like with Mollow triplet states, these build up only after a steady-state regime?} \kami{panels (a,b) are eigenenergies not quasienergies, so we are still talking about $t=0^+$ where the effect of $\eta_M$ is effective only but not the oscillation yet.}
% \sh{With finite $\omega_M$, they are certainly quasienergies, which is the only thing you are referring to with finite time of course; these do not make sense at small $t$, but come from cycle averaging, as far as I understand them}.

In Fig.~\ref{Fig2}(b), we show the eigenenergies of the 
{\it time-independent} part of the total Hamiltonian in Eq.~\eqref{HC_FQR},  characterized by $\mathcal{H}_0$ in the expansion terms in Eq.~\eqref{H_m}, in comparison with the eigenenergies of the QRM, versus  
$\eta_{ M}$, for $\eta_0=0$.
There is a shifting (renormalization) of the eigenenergies, and then 
a modified anticrossing 
%or the effective level splitting 
of the time-independent (static) eigenenergies.
%(as the result of the time dependence being turned on).
% This result 
% %particularly 
% emphasizes the impact of the nonperturbative approach, since in the perturbative approach, one constructs the time-dependent perturbation from the QRM Hamiltonian. 
We also
%particularly witness 
recognize that as one transitions toward the USC
(i.e., $\eta_M>0.1$), the shifts are amplified until they form new anticrossing regions. Note these shifts are independent of the 
$\omega_M$.
%
%time-dependent coupling frequency.
%[since they  relate to 
%$J_0[2(a+a^\dagger)\eta_M]$ in %Eq.~\eqref{H_m}].
%
When $\eta_0=\eta_{ M}=0$, there is no QRM dressing and no hybridization between the light and matter states [dashed green lines in panel~\ref{Fig2}(b)]. 
%and thus the states are the joint bare light and bare matter states. 
By increasing the static coupling $\eta_0>0$ when $\eta_{ M}=0$ [dashed lines in 
panel~\ref{Fig2}(a)], the light-matter hybridization begins, where the states are closer to each other, e.g., the first excited state is pushed closer to the ground state, and the second and third excited states  
move closer. 
%The effect of the renormalization shift,
%when $\eta_{ M}>0$, is that the hybridization becomes stronger. 
As $\eta_{ M}$ increases, this boosts the hybridization in such a way that they  form new anticrossings at certain field strengths,
which depends on the state levels. 

We next transform the problem 
%in
to the Floquet picture, including
the various $m$ solutions. 
In Fig.~\ref{Fig2}(c), the Floquet quasienergies within the $1^\mathrm{st}$ BZ are shown for the QRM truncated with sixteen states, so that at each value of the drive's parameters there exist (nominally) sixteen quasienergy states within one BZ.
%( we restrict this number here for clarity). 
Although the original initial-time states are the QRM states, the Floquet states are built upon the renormalized states 
from the dynamical coupling. 
% This 
% leads 
% %fact prompts 
% to two effects 
% %expectations 
% in the Floquet quasienergy diagram, shown
% in Fig.~\ref{Fig2}(c): 
The 
%addition of a 
DC component alters the
%restriction of the 
%dressed states 
transition selection rule between dressed states and induces significant intermixing of the QRM states. Due to the greater number of strongly coupled nearby states in the DC-renormalized Hamiltonian, nonlinear 
optical effects 
can 
%are expected to 
occur at a much lower dynamical coupling strength than they would be in the absence of the DC coupling,
%Thus, the introduction of an additional external DC component, %therefore, 
strongly enhancing the transition probabilities 
~\cite{Chu_Quantum_1982,Chu_Recent_1985,Chu_Beyond_2004}.
%and results in
%much richer spectral 
%features~\cite{Chu_Quantum_1982,Chu_Recent_1985,Chu_Beyond_2004}.
%[cf.~Fig.~\ref{Fig2}(a,b)].
% (i) the $l=0$ sideband line
% for any state $j$  roughly follows the renormalized line before any anticrossing occurs, i.e., the top thin bright line
% follows the thick red line and the bottom thin dark line follows the thick blue line, and (ii) 
This 
manifests in a rich Floquet quasienergy diagram, shown
in Fig.~\ref{Fig2}(c),
which 
%the emergence of the  anticrossing in the shifted eigenenergy [in Fig.~\ref{Fig2}(c)]  
yields 
%hints at 
%an early prediction of 
%in 
a large number of  
anticrossings~\cite{Ho_Floquet-Liouville_1798,Chu_Recent_1985,Breuer_Adiabatic_1989,Breuer_Quantum_1989,Chu_Beyond_2004,Holthaus_Floquet_2016}.
These quasienergies are continuous functions of the drive 
%strength 
amplitude that 
%do not cross, but
shows avoided crossings, if there are no symmetries that
allow 
%actual 
crossings.
%Such avoided crossings are abundant in the spectra of time-periodic systems.

%and are of central interest. 
%The influence of basis 
%size on the avoided crossings is %discussed in the SM~\cite{Akbari_Floquet_2023_SI}.

To help appreciate the 
dynamically-modified transitions, 
consider 
%the system is in
the QRM ground state 
%at $t=0^-$ or in $\vert j=0\rangle$
at $t=0^+$. As time evolves, with $\eta_{ M}\neq0$, the ground state adiabatically transforms into a Floquet state.
%evolving by the effect of the external stimulation with an energy shift. 
At the first anticrossing between a Floquet sideband of the lower state and a Floquet sideband of a higher state, where a superposition of the Floquet sidebands is constructed, a diabatic transition, namely, $\lvert\alpha_l\rangle\to\lvert\alpha'_{l'}\rangle$ at the anticrossing where a superposition $a\lvert\alpha_l\rangle+a'\lvert\alpha'_{l'}\rangle$ is provided.
%, which 
%is expected to occur to a higher Floquet state. 
Then, the created superposition, generally oscillating at a multiple of the stimulation frequency $\Delta\varepsilon=n\omega_{ M}$, adiabatically transforms into the superposition of the two original QRM states, and now oscillates at their energy difference, $E_{kj}=E_k-E_j$. 
Depending on the size of $\omega_{ M}$, one can go back and forth among different BZs to form a transition between the QRM states, if it is 
parity 
allowed.

%--- DISCUSSION OF FIG3 ---
In the USC regime, transitions are not between the bare
states of the system (with fixed numbers of photons and atomic excitations), but between the {\it dressed states of the composite system}.
%(which contain contributions from various numbers
%of photons and atomic excitations). 
%The Hamiltonian of such a system is non-number conserving, and its ground state contains a finite population of virtual excitations~\cite{DeLiberato_Virtual_2017,FriskKockum_Ultrastrong_2019}.
Indeed, naively assuming 
%that
the emitted radiation is  proportional to the photon population in the cavity 
%neglecting to discriminate between real and virtual particles 
leads to the prediction of unphysical radiation from the ground state in vacuum~\cite{Beaudoin_Dissipation_2011,Gustin_Gauge-invariant_2023,Salmon_Gauge-independent_2020,Mercurio_Flying_2022}. 
Instead, one must
use the correct dressed system 
operators~\cite{Settineri_Dissipation_2018,Salmon_Gauge-independent_2020}
\begin{equation}
%    \begin{split}
        s^{\Lambda +}=\sum_{j,k>j}S^{\Lambda}_{jk}\,\lvert j\rangle\langle k\rvert,
 %   \end{split}
    \label{}
\end{equation}
where $s^{\Lambda -}=[s^{\Lambda +}]^\dagger$, with $\Lambda=\{\text{cav},\text{TLS}\}$ and $S^{\Lambda}_{jk}\equiv\langle j\vert S^\Lambda\vert k\rangle$ is the matrix element of the system operator %at the initial time (or, 
in the Schr\"{o}dinger picture. 
Specifically, we
use $S^{\rm cav}
= a(1+\ii)/\sqrt{2} + {\rm H.c.}$~\cite{Hughes_Reconciling_2023},
and $S^{\rm TLS}=\sigma_x$.

Next, we study how one can 
produce real photons from the 
{\it time-dependent} perturbation. 
We assume that the system is initially in the dressed ground state $\lvert j=0\rangle$, and neglect thermal excitations. %effects.  
The 
number of 
real photon or  
%cavity (real) photon number or the number of the 
TLS excitations are defined from~\cite{Mercurio_Flying_2022,Salmon_Gauge-independent_2020,Settineri_Dissipation_2018}
%\begin{equation}
%    \begin{split}
        $N_{\Lambda}(t)=\langle\psi(t)\vert s^{\Lambda -}s^{\Lambda +}\vert\psi(t)\rangle$.
%    \end{split}
%    \label{Nt}
%\end{equation}
In contrast, 
the 
virtual photon number in the 
%(possibly entangled) 
ground state of the time-independent light-matter Hamiltonian 
is~\cite{Mercurio_Flying_2022,DeLiberato_Virtual_2017}
$\langle0\vert a^\dagger a\vert0\rangle_{\rm QR}$, which is  nonzero in vacuum USC~\cite{DeLiberato_Virtual_2017,FriskKockum_Ultrastrong_2019}. 
An observable, $\langle\psi(t)|O|\psi(t)\rangle$, is not necessarily time periodic due to the presence of off-diagonal terms
in the Floquet eigenbasis~\cite{Eckardt_Atomic_2017}, $\mathrm{e}^{\ii(\varepsilon_\alpha-\varepsilon_\beta)(t-t_0)}
\langle\alpha(t)|O|\beta(t)\rangle$, for $\alpha\neq\beta$
(see Fig.~S3 in~\cite{Akbari_Floquet_2023_SI}). However,
in real open systems,
 the off-diagonal terms are suppressed,
% there are various situations where the
% contributions of the off-diagonal terms are suppressed,
% such as in open systems. In those cases,
and the time evolution of observables
often becomes periodic with the same periodicity as the driving
field. 
We thus 
%A convenient option is to 
add a phenomenological damping rate, $\gamma$, to %only 
the non-diagonal terms,
which are then damped out 
after a sufficiently long time~\cite{Chu_Beyond_2004,DeLiberato_Virtual_2017,Eckardt_Atomic_2017}.
%so that the terms with $\alpha\neq\beta$ disappear after a sufficiently long time. 
Subsequently, we derive the expectation values,
\begin{equation}
\label{N}N_\Lambda(t)=\sum_{\alpha\beta}c^*_\alpha c_\beta\mathrm{e}^{\ii(\varepsilon_\alpha-\varepsilon_\beta)t-\gamma t(1-\delta_{\alpha\beta})}\langle\alpha(t)\lvert s^{\Lambda-}s^{\Lambda+}\rvert\beta(t)\rangle,
\end{equation}
and obtain the 
steady-state values: $N_\Lambda(t> t_{\rm ss})=\sum_{\alpha}\lvert c_\alpha\rvert^2\langle\alpha(t)\lvert s^{\Lambda-}s^{\Lambda+}\rvert\alpha(t)\rangle$.
 %We then define 
 The 
  mean real excitation number 
 is
 $ \overline{N}_{\Lambda}=\frac{1}{T}\int_{t_\mathrm{ss}}^{t_\mathrm{ss}+T} dt\,N_{\Lambda}(t)$, where $T$ is sufficiently long 
 %enough 
 to yield a steady-state average.
Our result is exactly $T$ periodic,
 so we only have to use one period. 

% We neglect model-specific loss interactions,
% %which have shown to
% however, these have been shown
% to only affect the predicted real photon numbers
% by a few percent for typical USC systems~\cite{DeLiberato_Virtual_2017}.

\begin{figure}[h]
    \centering
\includegraphics[width=.99\linewidth]{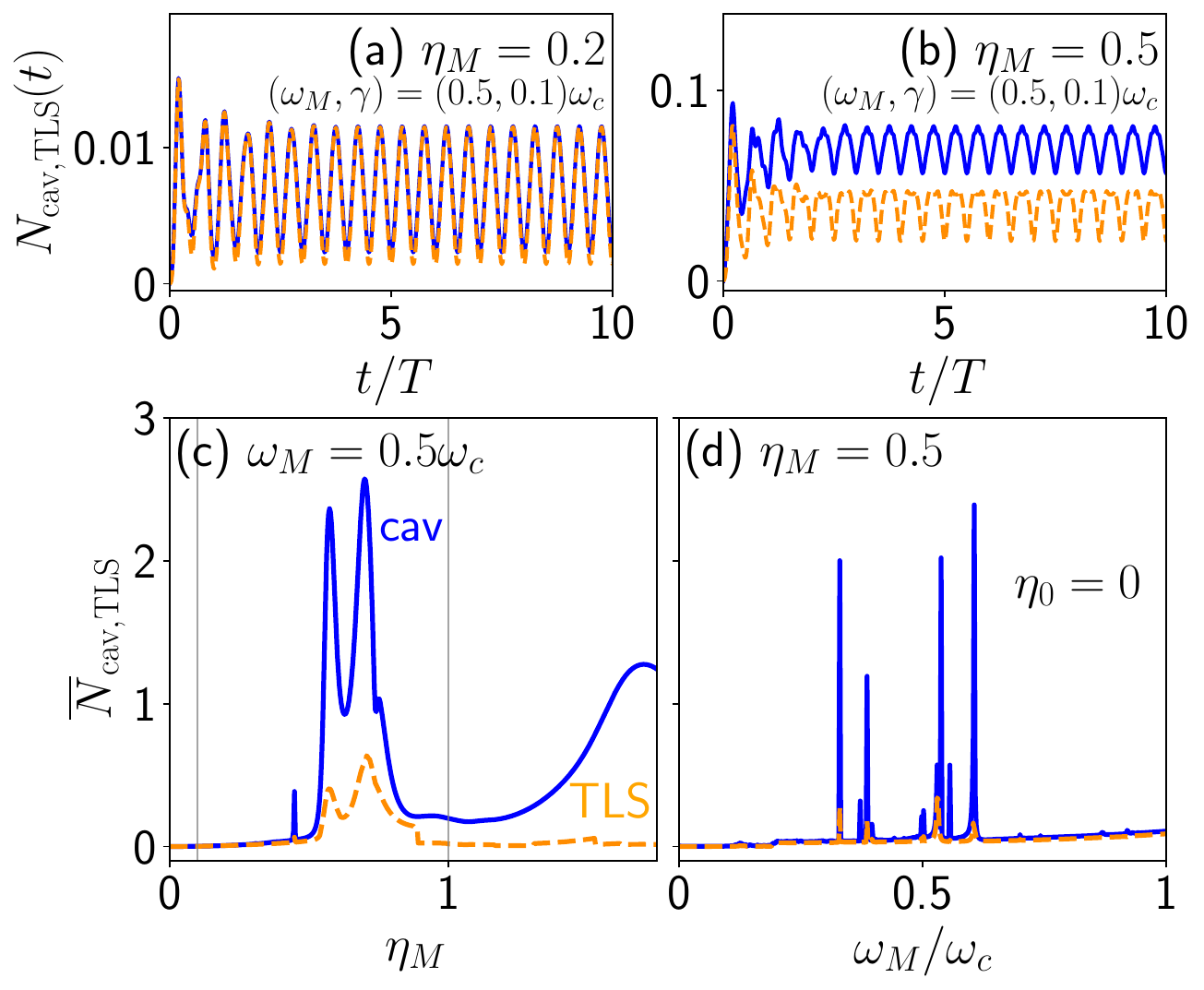}
\caption[]{
%\textbf{Cavity photon number.}
An example of the time-dependent real excitations, $N_{\rm cav}(t)$ (real photons, solid blue line) and $N_{\rm TLS}(t)$ (real TLS excitation, dashed orange line), given by Eq.~\eqref{N}, are shown for (a) $\eta_{M}=0.2$ and (b) $\eta_{M}=0.5$, with $\gamma=0.1\omega_c$. 
Also, depicted are the \emph{mean excitation numbers} that are the temporal average of the cavity excitation number, $\overline{N}_{\rm cav}$ (solid blue line) and the TLS excitation number, $\overline{N}_{\rm TLS}$ (dashed orange line), versus the amplitude (c) and frequency (d) of the dynamical coupling.
The parameters used are: $\eta_0=0$, $\omega_a=\omega_c$ and $m_{\rm max}=l_{\rm max} = 20$.
The grey vertical at $\eta_M=0.1$ and $\eta_M=1$ in panel (c) span the USC to deep-USC regime.
     }
    \label{Fig3}
\end{figure}

In Fig.~\ref{Fig3}, we show $N_{\rm cav,TLS}(t)$, with zero static coupling, $\eta_0=0$, with different dynamical values of (a,b) $\eta_{ 
 M}=0.2, 0.5$, 
 %and (b) $\eta_M = 0.5$, 
 using 
 %a  mechanical vibration frequency of
 $\omega_{M}=0.5\omega_c=0.5\omega_a$. 
 %For our purpose, we safely see $t_{\rm ss}\sim10T$.  
 We note that the onset and time range of the periodic behavior of the populations after they reach the steady state are different for different values of the coupling.
  %As expected, 
  The results are periodic after a sufficiently long time, depending on the strength and frequency of dynamical coupling;
  we  safely define
 $t_{\rm ss}\sim5T$.
 % For larger couplings, 
 % %as expeced, 
 %  need more time to settle and repeat.
For increasing coupling,
% we see that
the periodic modulation causes
a significant population of real photons (solid curves) and TLS excitation (dashed curves). This scenario requires
$\eta_{M}\neq0$ and the USC regime. 
Note 
%also 
that the populations
of the cavity are different to the TLS for increasing $\eta_M$, as the nonlinear behavior of the TLS and cavity photons are 
%quite 
different for increasing drives.

 % Figure 3 ---------------------------------------
%--------------------------------------------

In Fig.~\ref{Fig3}(c), we plot the average real excitation numbers,
versus
$\eta_M$,
with 
$\omega_M=0.5\omega_c$
and $\eta_0=0$.
 Finite  $\eta_0$
simulations are discussed in the supplement~\cite{Akbari_Floquet_2023_SI}.
We also observe that the onset of USC (or switching on/off within the USC, i.e., the joint effect of the static and dynamical couplings) coincides with the starting point of turning virtual photons into real ones, where there exists the discrepancy between real and virtual photons~\cite{Mercurio_Flying_2022}, i.e.,
$\eta_M\geq0.1$, since  $\eta_0=0$.
In Fig.~\ref{Fig3}(d), we plot the variation of the mean real excitation number versus $\omega_{M}$, with a fixed $\eta_M=0.5$, for the cavity (solid blue) and the TLS (dashed orange) excitations.
From panel~(d), we generally understand that
because the switching on/off process
is already in the USC, namely, because $\eta_M>0.1$, the starting point of turning virtual photons into real ones 
begins as soon as $\omega_M>0$, where we also observe a 
difference 
%discrepancy 
between the real photons and the TLS excitations,
%for the same reason as that discussed for Fig.~\ref{Fig3}(c). This discrepancy 
which is also higher when the amplitude and frequency of the mechanical motion are larger.

The results in Fig.~\ref{Fig3}(c,d) are obtained for sixteen QRM states and $m_{\rm max}=l_{\rm max}=20$. Adding more truncated QRM states may modify some of the frequency/coupling peaks,
and add additional
sharper peaks, but in practice these will be broadened
%further in experiments
with dissipation processes. 
Importantly, our main predictions are not qualitatively affected by a further increase
in basis size. 
The general intuitive behavior of the spectral shape is that as the amplitude and frequency of the dynamical coupling increase within the USC range, and the number of real photons and TLS excitations become larger 
%as well as the discrepancy between them 
because of the enhanced nonlinearity of the quantum processes.

The peak and valley structures
seen in Fig.~\ref{Fig3}(c,d) 
%(which are the signatures of the nonperturbative Floquet engineering)
are connected to the anticrossings of the Floquet quasienergy spectrum. 
Moreover, higher multi-oscillation peaks are narrower than lower-oscillation peaks and 
%also 
they form earlier (smaller values) in amplitude and frequency of the drive. This general effect of an 
increasing spectral width of an absorption line with the increase in the steady source intensity is similar to the power broadening effect in atomic absorption spectra~\cite{Vitanov_Power_2001}. 

To highlight some general features, the
major cavity double-peak and valley structure in Fig.~\ref{Fig3}(c) is actually formed by the combination of a $3$-$\omega_M$ resonance transition (from $\lvert j=0\rangle\to\lvert j=3\rangle$) and a $15$-$\omega_M$ resonance transition (from $\lvert j=0\rangle\to\lvert j=15\rangle$). 
%This structure is due to the transition from the ground state to the eleventh excited state of the renormalized DC Hamiltonian (see Fig.~S2(c) of~\cite{Akbari_Floquet_2023_SI}). 
% and has gradually undergone through dynamical AC-Stark shift as the number of oscillating harmonics increased. 
In the first BZ of the quasienergy diagram [Fig.~\ref{Fig1}(c) and S2(b)], the most effective corresponding anticrossing (which is quite wide as well due to nonlinear effects of power broadening) at the same points between 
%almost
the two Floquet sidebands $\lvert\alpha_l=12_l\rangle$ and $\lvert\alpha_l=14_l\rangle$ (see Fig.~S2 of supplement~\cite{Akbari_Floquet_2023_SI}). 
% By tracking these two sidebands,
% % in Fig.~S3(c), 
% it is evident that the former sideband translates to $\lvert\alpha'_{l'}=0_{l-1}\rangle$, where it is connected adiabatically to $\lvert j=0\rangle$ and the latter sideband translates to $\lvert\alpha'_{l'}=7_{l+10}\rangle$, to adiabatically construct the $\lvert j=11\rangle$ renormalized-QRM state. Hence, the transition is a $11$-$\omega_M$ event. 
% \kami{Moreover, analyzing the same point in Fig.~\ref{Fig:FQE_1}(b) points out that $E^{\rm renormalized}_{6,5}=E^{\rm renormalized}_{6}-E^{\rm renormalized}_{5}=\varepsilon_{6,5}+\omega_M=\varepsilon_{6}-\varepsilon_{5}+\omega_M$.}
% Similarly, by investigation and comparison of the plots in Figs.~\ref{Fig3} (and S2), 
% one can conclude that the other major peak at $\eta_M\approx 0.52$  in Fig.~\ref{Fig3}(c) is a $7$-$\omega_M$ peak coming from the excitation to the seventh renormalized QRM state. 
Furthermore, the very wide power-broadened peak at the far right of the panel is a $4$-$\omega_M$ peak due to the transition from the ground state to the fourth excited state. 

Note that not only the creation of each individual peak, but also the interplay between the different order transitions and peaks are crucial in the overall construction and understanding of the population spectra. These interplays include the constructive and destructive nonlinear interaction of peaks with various widths and strengths, which can cause a
 sudden dip or rise, and nonlinear effects~\cite{Autler_Stark_1955,Cohen-Tannoudji_Autler-Townes_1996,Ahmed_Autler-Townes_2012,Chu_Beyond_2004,Peng_What_2014} such as power broadening,
dynamical Stark shift, Autler–Townes multiplet splitting, electromagnetically induced transparency, and hole burning.
%, and S-hump behaviors, etc. 
For example, the drop at $\eta_M\sim0.9$ of the TLS graph in Fig.~\ref{Fig3}(c) is 
caused
%a clear-cut instance of 
by the destructive interfering of transitions. These modifications 
%in the spectrum 
arise due to Stark splitting of the driven system energy levels where the decaying system process (atomic down transition in the dressed states) from the two dressed states interferes destructively to create a Fano-type dark line in the single Lorentzian peak~\cite{Chu_Quantum_1982,Chu_Recent_1985,Cohen-Tannoudji_Autler-Townes_1996,Chu_Beyond_2004}.
Similar explanations are applicable for other resonant peaks in the same graph as well as those in Fig.~\ref{Fig3}(d),
which shows the role
of increasing
$\omega_M$ for fixed
$\eta_M=0.5$~\cite{Akbari_Floquet_2023_SI}.

% The rich peak structure
% highlights the higher-order multi-oscillation nonlinear quantum
% processes~\cite{Akbari_Floquet_2023_SI}. 

% For example, the peak at $\omega_M\approx0.6\omega_c$ of Fig.~\ref{Fig3}(d) is associated with a $9$-$\omega_M$ quantum process to transit the state of the system from the ground to the tenth excited state of the renormalized DC Hamiltonian, whereas we see a $4$-$\omega_M$ peak at about $0.53$ due to the transition from the ground to the forth excited state, a $7$-$\omega_M$ peak at about $0.5$ due to the transition from the ground to the seventh excited state, and a $17$-$\omega_M$ peak at about $0.33$ due to the transition from the ground to the tenth excited state.
% Many of the salient features in the spectral lineshapes may be
% qualitatively understood in terms of an analytical three- or four-level model and these predictions are in
% accord with the experimental observations~\cite{Ho_Floquet-Liouville_1986}.

Lastly, we comment on the
role of the $\eta(t)$ waveform.
%effect of the shape of the dynamical coupling waveform. 
With a pure harmonic $\eta(t)$ waveform, one can drive frequencies comparable to a few fractions of the photon frequency, as 
%largely discussed 
known in the context of the dynamical Casimir effect.
Generally, one understands that the production of virtual to real excitations must be done nonadiabatically, which means a sudden switch on and switch off of an interaction. Hence, it is expected that nonsmooth waveforms, such as a periodic array of sudden ramps (sawtooth) or top-hat functions, are comparatively highly productive~\cite{Ono_Quantum_2019}.
These waveforms should not depend too much on $\omega_M$, since the turn on and turn off is nonadiabatic, like how stimulated Raman adiabatic passage works best with a triangular pulse~\cite{Drese_Floquet_1999} (see~\cite{Akbari_Floquet_2023_SI}).

% \clearpage\newpage
% \kami{Calculate equation 5 and fig. 2b of the nanophotonics paper for real and virtual photons.
% $\langle s^{\mathrm{cav}(-)}(t) s^{\mathrm{cav}(+)}(t)\rangle= \langle s^{\mathrm{cav}(-)}(t) s^{\mathrm{cav}(+)}(t)\rangle$}

%%%%%%%%%%%%%%%%%%%%%%%%%%%%%%%%%%%%%%%%%%%%%%%%%%%%%%%%
% section: conclusion
%%%%%%%%%%%%%%%%%%%%%%%%%%%%%%%%%%%%%%%%%%%%%%%%%%%%%%%%
% \section{Conclusions}
% \label{Sec:Conclusions}
To conclude, we have investigated the dynamics of a cavity-QED system prepared initially in its lowest-dressed state 
subject to a time-dependent coupling rate in the USC regime, a Floquet engineered QRM. By using a suitable gauge-invariant model Hamiltonian, we 
described the generation of real excitations out of vacuum. This release of energy, which is initially stored in the form of virtual particles as a signature of the USC regime, is feasible by the nonadiabatic change in the system Hamiltonian, which is manifested
by the  mechanical oscillation of the location of the atom (or, also the cavity).
%(or, equivalently, the cavity walls).
%, which injects a classical driving field into the system.

This work was supported by the Natural Sciences and Engineering Research Council of Canada (NSERC), the National Research Council of Canada (NRC), the Canadian Foundation for Innovation (CFI), and Queen’s University, Canada. S.H. acknowledges the Japan Society for the Promotion of Science (JSPS) for funding support through an Invitational Fellowship. F.N. is supported in part by the Nippon Telegraph and Telephone Corporation (NTT) Research, the Japan Science and Technology Agency (JST) [via the Quantum Leap Flagship Program (Q-LEAP), and the Moonshot R\&D Grant Number JPMJMS2061], the Asian Office of Aerospace Research and Development (AOARD) (via Grant No. FA2386-20-1-4069), and the Office of Naval Research (ONR) Global (via Grant No. N62909-23-1-2074).

\acknowledgements
% This work was supported by the Natural Sciences and Engineering Research Council of Canada (NSERC),
% the National Research Council of Canada (NRC),
% the Canadian Foundation for Innovation (CFI), and Queen's University, Canada.

%%%%%%%%%%%%%%%%%%%%%%%%%%%%%%%%%%%%%%%%%%%%%%%%%%%%%%%%
% section: bibliography
%%%%%%%%%%%%%%%%%%%%%%%%%%%%%%%%%%%%%%%%%%%%%%%%%%%%%%%%
\bibliography{main}

% \clearpage
% \newpage
% %%%%%%%%%%%%%%%%%%%%%%%%%%%%%%%%%%%%%%%%%%%%%%%%%%%%%%%%
% % section: appendices
% %%%%%%%%%%%%%%%%%%%%%%%%%%%%%%%%%%%%%%%%%%%%%%%%%%%%%%%%
% \appendix

% \section{}
% \label{Sec:}

%%%%%%%%%%%%%%%%%%%%%%%%%%%%%%%%%%%%%%%%%%%%%%%%%%%%%%%%%%%%%%%%%%%%%%
%%%%%%%%%%%%%%%%%%%%%%%%%%%%%%%%%%%%%%%%%%%%%%%%%%%%%%%%%%%%%%%%%%%%%%
\end{document}